\documentclass[amsmath,amssymb,iopams]{iopart}
\usepackage{graphicx}

\begin{document}

%%%%%%%%%%%%%%%%%%%%%%%%%%%%%%%%%%%%%%%%%%%%%%%%%%%%%%%%%%%%
\title
{
The Peierls-Hubbard model at weak coupling
}
%%%%%%%%%%%%%%%%%%%%%%%%%%%%%%%%%%%%%%%%%%%%%%%%%%%%%%%%%%%%
\author{Michael Dzierzawa and Carmen Mocanu}
\address{Institut f\"ur Physik, Universit\"at Augsburg, 86135 Augsburg, Germany}
\ead{michael.dzierzawa@physik.uni-augsburg.de}
\date{\today}

\begin{abstract}
We investigate the Peierls transition in the one-dimensional Peierls-Hubbard
model at half filling in the adiabatic approximation for the lattice.
Depending on the value of the
electron-lattice coupling constant $g$ the equilibrium dimerization can
be either enhanced or suppressed by the Hubbard interaction $U$.
Applying second order perturbation theory
we determine the critical value $g_c = 0.689348$ below which the
Hubbard interaction enhances the dimerization.

\end{abstract}


\section{Introduction}
%%%%%%%%%%%%%%%%%%%%%%%%%%%%%%%%%%%%%%%%%%%%%%%%%%%%%%%%%%%%
The coupling of a one-dimensional metal to an elastic lattice results in an instability
towards a lattice distortion known as Peierls transition \cite{Peierls55}.
This phenomenon can be observed in various
quasi-one-dimensional materials, e.g.\ conjugated polymers \cite{Keiss92} like polyacetylene,
charge-transfer salts \cite{Ishiguro90}, or spin-Peierls compounds like ${\rm Cu Ge O_3}$ .

From a theoretical point of view, a first step
towards a quantitative description of the Peierls transition
has been made by Su, Schrieffer and Heeger (SSH) \cite{Su80}.
However, in their model neither the dynamics of the lattice nor the electron-electron
interaction has been taken into account. Subsequently there has been made some effort
to include phonons beyond the adiabatic approximation.
While in  the approach of SSH the Peierls transition occurs
for any nonzero value of the electron-phonon (el-ph) coupling,
a minimum strength of the el-ph coupling is required for the transition
if quantum phonons are included \cite{Sengupta03}.

In this paper we consider the limit of a static lattice and focus on the role of
the electron-electron interaction.
Correlation effects on the Peierls transition have been studied using a variety
of methods including variational wave-functions \cite{Horsch81,Baeriswyl85}, Hartree Fock
plus perturbation theory \cite{Kivelson82},
quantum Monte Carlo \cite{Hirsch83}, numerical diagonalization of small
systems \cite{Mazumdar83,Hayden88,Waas90,Ogata93}, bosonization \cite{Sugiura02,Mocanu04}, and
incremental expansion \cite{Malek98,Malek03}.
It is now generally accepted that
for small values of the el-ph coupling constant $g < g_c$
the dimerization is enhanced when the Hubbard interaction $U$ is switched on
while for $g > g_c$ it is suppressed.
Estimates for the critical el-ph coupling that have been obtained in the literature
are $g_c \approx 0.75$ in \cite{Baeriswyl85},
$g_c \approx 0.76$ in \cite{Waas90} and $g_c \approx 0.69$ in \cite{Malek98,Malek03}.
For $g$ not too close to $g_c$ the dimerization reaches a maximum at $U$ of the order of the bandwidth
and goes to zero for $U \rightarrow \infty$.

Apart from this qualitative agreement there is still a lack of rigorous results,
in particular in the limit of weak el-ph coupling $g \ll g_c$
which is difficult to address in the aforementioned approaches.
In order to remove this uncertainties we
study the Peierls-Hubbard model in the weak coupling regime
using second order perturbation theory with respect to the
Hubbard interaction. This allows use to derive rigorous results in the
limit $U \rightarrow 0$ for all values of the el-ph coupling.

The paper is organized as follows:
In Section two we introduce the Peierls-Hubbard model and discuss the
validity of the perturbative approach. The main results concerning the Peierls transition
are presented in Section three.
In Section four we discuss our results and compare them with different approaches.

%%%%%%%%%%%%%%%%%%%%%%%%%%%%%%%%%%%%%%%%%%%%%%%%%%%%%%%%%%%%
\section{Model and perturbation theory}
%%%%%%%%%%%%%%%%%%%%%%%%%%%%%%%%%%%%%%%%%%%%%%%%%%%%%%%%%%%%
%\subsection{The model}
%%%%%%%%%%%%%%%%%%%%%%%%%%%%%%%%%%%%%%%%%%%%%%%%%%%%%%%%%%%%
We consider the one-dimensional Peierls-Hubbard model
%%%%%%%%%%%%%%%%%%%%%%%%%%%%%%%%%%%%%%%%%%%%%%%%%%%%%%%%%%%%
%\begin{eqnarray}
%H&=& - \sum_{i,\sigma} (t - \alpha(u_{i+1} - u_i)) (c_{i\sigma}^\dag c_{i+1,\sigma} + {\rm h.c.})\\
%% & & - \sum_{i,\sigma} (t' + \alpha'(u_{i+2} - u_i))(c_{i\sigma}^\dag c_{i+2,\sigma} + {\rm h.c.}) \\
% & & + U \sum_{i} n_{i\uparrow} n_{i\downarrow} + \frac{K}{2} \sum_i (u_{i+1} - u_i)^2
%\label{Hamilt}
%\end{eqnarray}
%%%%%%%%%%%%%%%%%%%%%%%%%%%%%%%%%%%%%%%%%%%%%%%%%%%%%%%%%%%%
\begin{eqnarray}
\label{Hamilt}
H & = & - t\sum_{i,\sigma} (1 - \alpha (u_{i+1} - u_i))(c_{i\sigma}^\dag c_{i+1,\sigma} + {\rm h.c.})
  \nonumber\\
  & & + U \sum_{i} n_{i\uparrow} n_{i\downarrow} + \frac{K}{2} \sum_i (u_i - u_{i+1})^2
%- t\sum_{i,\sigma} (1 + (-)^i \delta) (c_{i\sigma}^\dag c_{i+1,\sigma} + {\rm h.c.})\\
% + 2 K N \delta^2
\end{eqnarray}
%%%%%%%%%%%%%%%%%%%%%%%%%%%%%%%%%%%%%%%%%%%%%%%%%%%%%%%%%%%%
where $c_{i\sigma}^\dag (c_{i\sigma})$ creates (annihilates) an electron
with spin $\sigma(=\uparrow, \downarrow)$ on a lattice of $N$ sites,
$n_{i\sigma} = c_{i\sigma}^\dag c_{i\sigma}$,
and $u_i$ is the deviation of the $i$'th atom from its equilibrium position in units
of the lattice constant $a$.
$U$ and $K$ are the on-site Hubbard interaction and the elastic constant
of the lattice, respectively, and $t$ is the hopping parameter.
In (\ref{Hamilt}) the phonons are treated in the adiabatic limit which is formally
obtained by letting the mass of the atoms go to infinity.
Since the lattice distortion is assumed small compared to
the lattice constant, the modification of the hopping amplitude can be
treated in linear order via the parameter $\alpha$.
In the following we restrict ourselves to half filling where due to $2 k_F = \pi/a$
the Peierls instability leads to an alternating lattice distortion
%%%%%%%%%%%%%%%%%%%%%%%%%%%%%%%%%%%%%%%%%%%%%%%%%%%%%%%%%%%%
\begin{equation}
\label{uofi}
u_i = (-1)^i \, u
\end{equation}
%%%%%%%%%%%%%%%%%%%%%%%%%%%%%%%%%%%%%%%%%%%%%%%%%%%%%%%%%%%%
It is convenient to express the el-ph coupling and the dimerization amplitude
by the dimensionless parameters $g$ and $\delta$, respectively,
which are defined by
%%%%%%%%%%%%%%%%%%%%%%%%%%%%%%%%%%%%%%%%%%%%%%%%%%%%%%%%%%%%
\begin{eqnarray}
\label{coupling}
g & = & \alpha \sqrt{\frac{t}{K}} \\
\delta & = & 2 \alpha u
\end{eqnarray}
%%%%%%%%%%%%%%%%%%%%%%%%%%%%%%%%%%%%%%%%%%%%%%%%%%%%%%%%%%%%
In terms of these parameters Hamiltonian (\ref{Hamilt}) reads
%%%%%%%%%%%%%%%%%%%%%%%%%%%%%%%%%%%%%%%%%%%%%%%%%%%%%%%%%%%%
\begin{equation}
\label{Hamilt1}
H = - t\sum_{i,\sigma} (1 + (-1)^i \delta) (c_{i\sigma}^\dag c_{i+1,\sigma} + {\rm h.c.})
+ U \sum_{i} n_{i\uparrow} n_{i\downarrow} + \frac{N t \delta^2}{2g^2}
\end{equation}
%%%%%%%%%%%%%%%%%%%%%%%%%%%%%%%%%%%%%%%%%%%%%%%%%%%%%%%%%%%%
We consider $t$, $U$ and $g$ as given model parameters while the dimerization $\delta$
is free to adjust itself such that the total energy is minimal.
In the following we measure all energies in units of the
hopping amplitude $t$ (i.e.\ we set $t = 1$).
%%%%%%%%%%%%%%%%%%%%%%%%%%%%%%%%%%%%%%%%%%%%%%%%%%%%%%%%%%%%
%\subsection{Perturbation theory}
%%%%%%%%%%%%%%%%%%%%%%%%%%%%%%%%%%%%%%%%%%%%%%%%%%%%%%%%%%%%
The ground state energy of the Hamiltonian (\ref{Hamilt1}) can only be calculated analytically
in the limiting cases $\delta = 0$ and $\delta = 1$, respectively.
$\delta = 0$ corresponds to the ordinary Hubbard model which has been solved using Bethe
ansatz \cite{Lieb68} and for $\delta = 1$ the system is completely dimerized.
For other values of $\delta$ one has to resort to approximations or to numerical methods.
Since we are only interested in the weak coupling regime we use second order
perturbation theory with respect to the Hubbard interaction. First of all
we want to justify the assertion that perturbation theory can indeed be applied for the
model (\ref{Hamilt1}).
In the case $\delta = 0$ the exact ground state energy known from the Bethe ansatz solution
\cite{Lieb68} can be expressed as an integral over Bessel functions.
As pointed out by Economou and Poulopoulos \cite{Economou79}
this integral can be converted into an asymptotic series in $U$.
Metzner and Vollhardt \cite{Metzner89} showed
that the exact second order term of this asymptotic expansion
can be calculated using perturbation theory.
Furthermore, they conjectured that this is true
not only for the second order term but for all coefficients of the expansion.
For $\delta > 0$ the conditions for applying perturbation theory are more favorable
than for $\delta = 0$ since the unperturbed ground state energy $E_0$
is separated from all excited states by an energy gap
$\Delta = 4\delta$ which avoids divergent contributions from small denominators.
We expect therefore that the ground state energy of the Peierls-Hubbard model
can be expanded in a series with a finite radius of convergence, $U_0(\delta)$
for all $\delta > 0$ and that the coefficients of this expansion can be obtained from
perturbation theory.
Krivnov and Ovchinnikov \cite{Krivnov84} have pointed out that within
the parquet approximation for a continuum version of the
model (\ref{Hamilt1}) the leading n'th order contribution to the ground
state energy is $\propto \delta^2 U^n \ln^{n+1} \delta$. Similar
conclusions were obtained by Horovitz and Solyom \cite{Horovitz85} using
a scaling approach and by Kivelson {\it et al} \cite{Kivelson86} by
comparison of exact results on the massive Thirring model
with the perturbation theory for a model of spinless
fermions. Fermions with spin were considered in \cite{Wu86} and again
a perturbation expansion in powers of $U \ln \delta$ was
obtained. Thus the radius of convergence behaves as $U_0(\delta)
\propto 1/|\ln \delta|$ and even for exponentially small
dimerization $\delta$ perturbation theory can be applied
over some finite range of $U$.
Writing $\epsilon(\delta,U) = E(\delta,U)/N$, the ground state energy per site of the
electronic part of (\ref{Hamilt1}), the expansion of $\epsilon$ in powers of $U$ takes the form
%%%%%%%%%%%%%%%%%%%%%%%%%%%%%%%%%%%%%%%%%%%%%%%%%%%%%%%%%%%%
\begin{equation}
\epsilon(\delta,U) - \frac{U}{4} = \epsilon_0(\delta) +
 \epsilon_2(\delta)U^2 + \epsilon_4(\delta)U^4 + \ldots
\label{series}
\end{equation}
%%%%%%%%%%%%%%%%%%%%%%%%%%%%%%%%%%%%%%%%%%%%%%%%%%%%%%%%%%%%
where we have used that due to particle-hole symmetry $\epsilon(\delta,U) - U/4$
is an even function of $U$.
In particular, in the case of complete dimerization,
$\delta=1$, it is straightforward to calculate the exact ground state energy
and expand it in a series which converges for $U < 8$:
%%%%%%%%%%%%%%%%%%%%%%%%%%%%%%%%%%%%%%%%%%%%%%%%%%%%%%%%%%%%
\begin{equation}
\label{epsu1}
\epsilon(1,U) - \frac{U}{4} = - 2 \sqrt{1 + \frac{U^2}{64}}
  = - 2 - \frac{U^2}{64} + \frac{U^4}{16384} + \ldots \;\; .
\end{equation}
%%%%%%%%%%%%%%%%%%%%%%%%%%%%%%%%%%%%%%%%%%%%%%%%%%%%%%%%%%%%
%%%%%%%%%%%%%%%%%%%%%%%%%%%%%%%%%%%%%%%%%%%%%%%%%%%%%%%%%%%%
\section{Equilibrium dimerization}
%%%%%%%%%%%%%%%%%%%%%%%%%%%%%%%%%%%%%%%%%%%%%%%%%%%%%%%%%%%%
In order to determine the equilibrium dimerization we
minimize the ground state energy of the Hamiltonian (\ref{Hamilt1}), including the lattice contribution, with
respect to $\delta$ and obtain the condition
%%%%%%%%%%%%%%%%%%%%%%%%%%%%%%%%%%%%%%%%%%%%%%%%%%%%%%%%%%%%
\begin{equation}
\epsilon_0'(\delta) + \epsilon_2'(\delta)U^2 + \epsilon_4'(\delta)U^4 + \ldots + \frac{\delta}{g^2} = 0
\label{mini}
\end{equation}
%%%%%%%%%%%%%%%%%%%%%%%%%%%%%%%%%%%%%%%%%%%%%%%%%%%%%%%%%%%%
Eq.\ (\ref{mini}) defines the function $\delta(U)$ which can as well be expanded in a series
containing only even powers of $U$
%%%%%%%%%%%%%%%%%%%%%%%%%%%%%%%%%%%%%%%%%%%%%%%%%%%%%%%%%%%%
\begin{equation}
\label{deltaofU}
\delta(U) = \delta_0 + \delta_2 U^2 + \delta_4 U^4 + \ldots
\end{equation}
%%%%%%%%%%%%%%%%%%%%%%%%%%%%%%%%%%%%%%%%%%%%%%%%%%%%%%%%%%%%
Reinserting  Eq.\ (\ref{deltaofU}) into Eq.\ (\ref{mini}) and comparing order by order in $U$
yields
%%%%%%%%%%%%%%%%%%%%%%%%%%%%%%%%%%%%%%%%%%%%%%%%%%%%%%%%%%%%
\begin{eqnarray}
\label{delta0}
\delta_0 & = & - g^2 \epsilon_0'(\delta_0) \\
\label{delta2}
\delta_2 & = & - \frac{\epsilon_2'(\delta_0)}{g^{-2} + \epsilon_0''(\delta_0)} \\
\label{delta4}
\delta_4 & = & - \frac{\frac{1}{2} \delta_2^2 \epsilon_0'''(\delta_0)
+ \delta_2 \epsilon_2''(\delta_0) + \epsilon_4'(\delta_0)}{g^{-2} + \epsilon_0''(\delta_0)}
\end{eqnarray}
%%%%%%%%%%%%%%%%%%%%%%%%%%%%%%%%%%%%%%%%%%%%%%%%%%%%%%%%%%%%
The zeroth order dimerization has been calculated in \cite{Su80}.
In the limit $g \ll 1$ it is given by
%%%%%%%%%%%%%%%%%%%%%%%%%%%%%%%%%%%%%%%%%%%%%%%%%%%%%%%%%%%%
\begin{equation}
\label{delta0app}
\delta_0 = 4 \exp\left(-\frac{\pi}{4g^2} - 1\right)
\end{equation}
%%%%%%%%%%%%%%%%%%%%%%%%%%%%%%%%%%%%%%%%%%%%%%%%%%%%%%%%%%%%
%\subsection{Second order}
%%%%%%%%%%%%%%%%%%%%%%%%%%%%%%%%%%%%%%%%%%%%%%%%%%%%%%%%%%%%
\begin{figure}[t]
\centerline{\includegraphics[height=6cm,width=9cm]{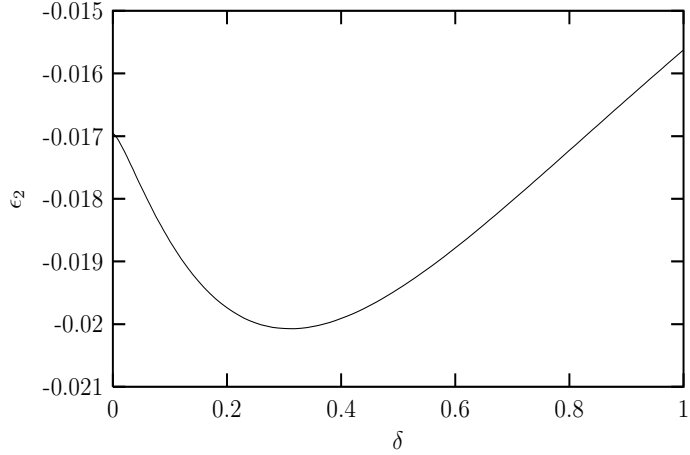}}
\caption{Second order contribution to the ground state energy per site,
$\epsilon = \epsilon_0 + \epsilon_2 U^2 + \epsilon_4 U^4 + \ldots$,
as function of the dimerization parameter $\delta$. The minimum occurs at
$\delta_c = 0.310523$.}
\label{fig:eps2}       % Give a unique label
\end{figure}
%%%%%%%%%%%%%%%%%%%%%%%%%%%%%%%%%%%%%%%%%%%%%%%%%%%%%%%%%%%%
Now we turn to the second order contribution $\propto U^2$ of the ground state energy.
Considering the Hubbard interaction in Eq. (\ref{Hamilt1}) as perturbation,
the second order coefficient of the series (\ref{series}) reads
%%%%%%%%%%%%%%%%%%%%%%%%%%%%%%%%%%%%%%%%%%%%%%%%%%%%%%%%%%%%
\begin{equation}
\epsilon_2(\delta) = - \frac{1}{N}\sum_{n \ne 0} \frac{|\langle n |D| 0 \rangle |^2}{E_n - E_0}
\label{eps2}
\end{equation}
%%%%%%%%%%%%%%%%%%%%%%%%%%%%%%%%%%%%%%%%%%%%%%%%%%%%%%%%%%%%
where $E_n$ is the energy of the eigenstate $|n \rangle$ of the noninteracting Hamiltonian $H_0$,
and $D = \sum_{i} n_{i\uparrow} n_{i\downarrow}$.
Representing $|n \rangle$ in the one-particle basis of $H_0$,
Eq.\ (\ref{eps2}) can be converted into
%%%%%%%%%%%%%%%%%%%%%%%%%%%%%%%%%%
\begin{equation}
\label{eps2int}
\epsilon_2(\delta) = - \frac{1}{32}
\int_{-\pi}^{\pi} \frac{{\rm d} k}{2\pi} \int_{-\pi}^{\pi} \frac{{\rm d} k'}{2\pi}
\int_{-\pi}^{\pi} \frac{{\rm d} q}{2\pi} \
%\frac{\left|1 + {\rm e}^{i(\varphi_{k+q}-\varphi_{k}+\varphi_{k'-q}-\varphi_{k'})} \right|^2}
%{|\Delta_{k+q}| + |\Delta_{k}| + |\Delta_{k'-q}| + |\Delta_{k'}|}
\frac{\left|1 + {\rm e}^{i(\varphi_{k+q}-\varphi_{k}+\varphi_{k'-q}-\varphi_{k'})} \right|^2}
{\varepsilon_{k+q} + \varepsilon_{k} + \varepsilon_{k'-q} + \varepsilon_{k'}}
\end{equation}
%%%%%%%%%%%%%%%%%%%%%%%%%%%%%%%%%%
where
%all integrals run from $-\pi$ to $\pi$ and
%%%%%%%%%%%%%%%%%%%%%%%%%%%%%%%%%%
\begin{eqnarray}
%\Delta_{k} = 1 + {\rm e}^{-ik} + \delta(1 - {\rm e}^{-ik}) = |\Delta_{k}|  {\rm e}^{i \varphi_k}
%\varepsilon_{k} {\rm e}^{i \varphi_k} = 1 + {\rm e}^{-ik} + \delta(1 - {\rm e}^{-ik})
\varepsilon_{k} & = & 2\sqrt{\cos^2 k + \delta^2 \sin^2 k} \\
{\rm e}^{i \varphi_{k}} & = & \frac{1 + {\rm e}^{-2ik} + \delta(1 - {\rm e}^{-2ik})}
{|1 + {\rm e}^{-2ik} + \delta(1 - {\rm e}^{-2ik})|} \;\;\; .
\end{eqnarray}
%%%%%%%%%%%%%%%%%%%%%%%%%%%%%%%%%%%%%%%%%%%%%%%%%%%%%%%%%%%%
Eq. (\ref{eps2int}) is equivalent with Eq.~(C5) in Ref. \cite{Kivelson82} in the
limit of zero staggered magnetization.
The integral (\ref{eps2int}) can be calculated numerically with very high precision.
The result shown in Fig.\ \ref{fig:eps2}
agrees with the analytical values $\epsilon_2(0) = -(7/16\pi^3) \zeta(3) = - 0.016961$
and $\epsilon_2(1) = -1/64 = -0.015625$. More importantly,
$\epsilon_2(\delta)$ has a minimum at $\delta_c = 0.310523$
%$\delta_c = 0.31052262$
which means that according to Eq.\ (\ref{delta2}) the sign of the coefficient
$\delta_2$ changes from plus to minus.
According to Eq.\ (\ref{delta0}) the critical dimerization $\delta_c$ corresponds to a
critical el-ph coupling constant
%%%%%%%%%%%%%%%%%%%%%%%%%%%%%%%%%%%%%%%%%%%%%%%%%%%%%%%%%%%%
\begin{equation}
\label{gc}
g_c = \sqrt{\frac{\delta_c}{|\epsilon_0'(\delta_c)|}} = 0.689348
\end{equation}
%%%%%%%%%%%%%%%%%%%%%%%%%%%%%%%%%%%%%%%%%%%%%%%%%%%%%%%%%%%%
Therefore for $g < g_c$ the dimerization is
enhanced by the Hubbard interaction while for $g > g_c$
it is suppressed.
%%%%%%%%%%%%%%%%%%%%%%%%%%%%%%%%%%%%%%%%%%%%%%%%%%%%%%%%%%%%
\begin{figure}[t]
\centerline{\includegraphics[height=6cm,width=9cm]{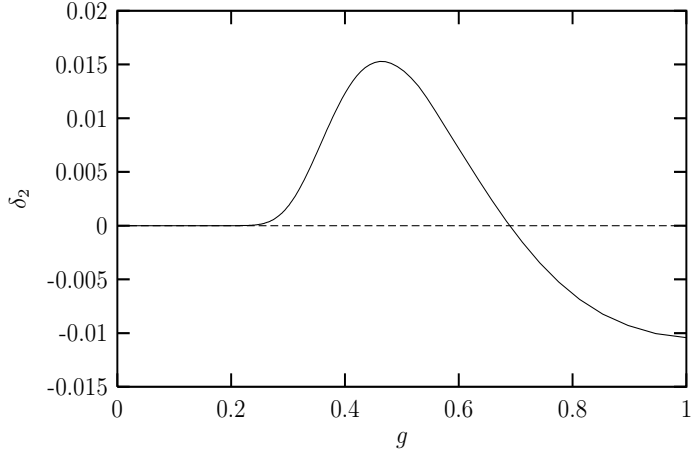}}
\caption{Second order coefficient $\delta_2$ of the equilibrium
dimerization  $\delta(U) = \delta_0 + \delta_2 U^2 + \delta_4 U^4 + \dots$
as function of the el-ph coupling constant $g$. Note that $\delta_2$ changes
sign at $g_c = 0.689348$ which is related to $\delta_c = 0.310523$ via Eq.\ (\ref{gc}).}
\label{fig:delta2}       % Give a unique label
\end{figure}
%%%%%%%%%%%%%%%%%%%%%%%%%%%%%%%%%%%%%%%%%%%%%%%%%%%%%%%%%%%%
The second order coefficient $\delta_2$ of Eq.\ (\ref{deltaofU}) which is displayed in Fig.\ \ref{fig:delta2}.
has a maximum at $g \approx 0.464$.
%which is not far away
%from the estimated value $g \approx 0.4$ for polyacetylene \cite{Baeriswyl85} which indicates
%that correlation effects cannot be ignored in this case.
In the limit of small el-ph coupling, $g \ll 1$,
Eq. (\ref{delta2}) simplifies to
%%%%%%%%%%%%%%%%%%%%%%%%%%%%%%%%%%%%%%%%%%%%%%%%%%%%%%%%%%%%
\begin{equation}
\label{delta2app}
\delta_2 = - \frac{\pi}{4}\epsilon_2'(\delta_0)
\end{equation}
From a careful analysis of the numerical data for small values of
$\delta$ we conjecture that
%%%%%%%%%%%%%%%%%%%%%%%%%%%%%%%%%%%%%%%%%%%%%%%%%%%%%%%%%%%%
\begin{equation}
\label{eps2app}
\epsilon_2(\delta) - \epsilon_2(0) \propto \delta^2 \left(\ln \frac{c}{\delta}\right)^3
\end{equation}
%%%%%%%%%%%%%%%%%%%%%%%%%%%%%%%%%%%%%%%%%%%%%%%%%%%%%%%%%%%%
with $c \approx 1.5$, in agreement with the result of the parquet summation \cite{Krivnov84}.
 Inserting this expression
into Eq. (\ref{delta2app}) and using
the asymptotic formula for $\delta_0(g)$ given in Eq. (\ref{delta0app}) we
obtain the result
%%%%%%%%%%%%%%%%%%%%%%%%%%%%%%%%%%%%%%%%%%%%%%%%%%%%%%%%%%%%
\begin{equation}
\label{delta2ofg}
\frac{\delta_2}{\delta_0} \propto g^{-6}
\end{equation}
%%%%%%%%%%%%%%%%%%%%%%%%%%%%%%%%%%%%%%%%%%%%%%%%%%%%%%%%%%%%
i.e.\ the relative weight of the second order term in the expansion
of the dimerization $\delta(U) = \delta_0 + \delta_2 U^2 + \delta_4 U^4 + \dots$
increases strongly for $g \rightarrow 0$, although both $\delta_0$ and $\delta_2$ are
exponentially small in this limit.
%%%%%%%%%%%%%%%%%%%%%%%%%%%%%%%%%%%%%%%%%%%%%%%%%%%%%%%%%%%%
%\subsection{Fourth order}
%%%%%%%%%%%%%%%%%%%%%%%%%%%%%%%%%%%%%%%%%%%%%%%%%%%%%%%%%%%%

\begin{figure}[t]
\centerline{\includegraphics[height=6cm,width=9cm]{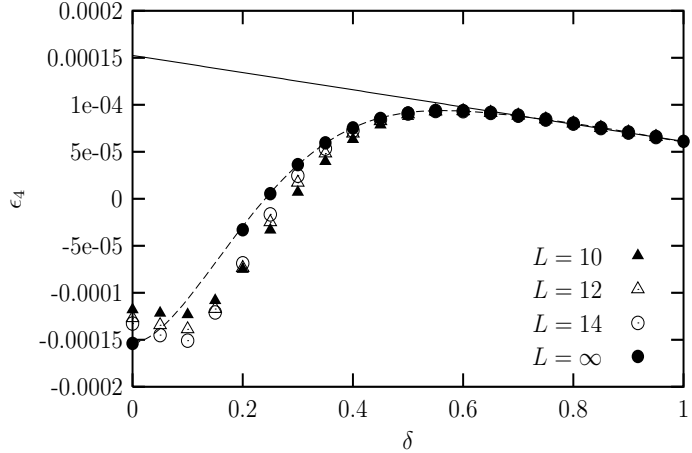}}
\caption{Fourth order contribution of the ground state energy per site,
$\epsilon(L) = \epsilon_0(L) + \epsilon_2(L) U^2 + \epsilon_4(L) U^4 + \ldots$,
as function of the dimerization parameter $\delta$ for $L = 10, 12, 14$ and $\infty$.
The dashed curve is meant as a guide to the eye, only. The solid line is
the asymptotic result $\epsilon_4(\delta) = 2^{-14}(1 + 3(1-\delta)/2)$
valid in the limit $\delta \rightarrow 1$.}
\label{fig:eps4}       % Give a unique label
\end{figure}

Unfortunately it becomes rather complicated to calculate the fourth order
term $\epsilon_4(\delta)$ using perturbation theory.
We have therefore chosen a different approach based on the exact numerical
diagonalization of small systems (up to $L = 14$ lattice sites) using the
Lanczos algorithm which allows us to calculate the ground state energy with
a relative precision of better
than $10^{-14}$. In order to avoid even-odd oscillations
we have implemented antiperiodic boundary conditions for $L = 4n$ and
periodic boundary conditions for $L = 4n + 2$. This guarantees that the
ground state is non-degenerate.
Since we know the exact second order
term $\epsilon_2(\delta,L)$ (for given system size $L$) it is possible to
extract $\epsilon_4(\delta,L)$ with very high precision.
However, one is left with the problem of extrapolating these values
to $L = \infty$. We have tried various extrapolation procedures.
For $\delta \ge 0.2$ the ansatz
$\epsilon_4(\delta,L) = \epsilon_4(\delta,\infty) + a \exp(-bL + c/L)$,
which is motivated by the existence of a gap in the excitation spectrum,
yields very stable results for $\epsilon_4(\delta,\infty)$.
In  particular, the asymptotic behavior close to the fully dimerized
limit, $\epsilon_4(\delta) = 2^{-14}(1 + 3(1-\delta)/2 + \ldots)$ for $\delta\rightarrow 1$
is correctly recovered (see Fig.\ \ref{fig:eps4}). However,
for $\delta < 0.2$ we could not perform a reliable
finite-size scaling of our data.
The reason is that the correlation length, which is inversely proportional to the
gap, is getting much larger than the size of the
systems that can be diagonalized numerically.
On the other hand, for $\delta = 0$ where there is no gap, an
unbiased polynomial fit yields
$\epsilon_4(0,\infty) = -1.547\times 10^{-4}$ which is quite close
to the exact value \cite{Economou79} $\epsilon_4(0) = - (93/2048\pi^5) \zeta(5)
= -1.538\times 10^{-4}$.
In Fig.\ \ref{fig:eps4} we plot $\epsilon_4(0,L)$ for $L = 10,12,14$ together with
the extrapolated value $\epsilon_4(0,\infty)$ (whenever available)
as a function of $\delta$. Comparing with the corresponding results for $\epsilon_2$
(see Fig.\ \ref{fig:eps2}) we see that $\epsilon_4$ is typically two orders
of magnitude smaller than $\epsilon_2$, which indicates that the weak coupling
approach should be quite accurate in the region $U \stackrel{<}{\sim} 1$.
%%%%%%%%%%%%%%%%%%%%%%%%%%%%%%%%%%%%%%%%%%%%%%%%%%%%%%%%%%%%
%%%%%%%%%%%%%%%%%%%%%%%%%%%%%%%%%%%%%%%%%%%%%%%%%%%%%%%%%%%%

For $g < g_c$ the dimerization $\delta(U)$ reaches a maximum at $U = U_{\rm max}$
which is given by
%%%%%%%%%%%%%%%%%%%%%%%%%%%%%%%%%%%%%%%%%%%%%%%%%%%%%%%%%%%%
\begin{equation}
U_{\rm max} = \sqrt{-\frac{\delta_2}{2\delta_4}}
\label{Umax}
\end{equation}
%%%%%%%%%%%%%%%%%%%%%%%%%%%%%%%%%%%%%%%%%%%%%%%%%%%%%%%%%%%%
when we neglect the terms of order $\sim U^6$ and higher in Eq.\ (\ref{deltaofU}).
Expanding $U_{\rm max}^2$ in powers of $g_c - g$ we obtain for $g$ close to $g_c$
%%%%%%%%%%%%%%%%%%%%%%%%%%%%%%%%%%%%%%%%%%%%%%%%%%%%%%%%%%%%
\begin{equation}
U_{\rm max}(g) \simeq \kappa (g_c - g)^{1/2}
\label{Umaxofg}
\end{equation}
%%%%%%%%%%%%%%%%%%%%%%%%%%%%%%%%%%%%%%%%%%%%%%%%%%%%%%%%%%%%
with
%%%%%%%%%%%%%%%%%%%%%%%%%%%%%%%%%%%%%%%%%%%%%%%%%%%%%%%%%%%%
\begin{equation}
\kappa = \frac{\delta_c \epsilon_2''(\delta_c)}
{g_c  \epsilon_4'(\delta_c) (1+g_c^2 \epsilon_0''(\delta_c))}
 \approx 8.7
\end{equation}
%%%%%%%%%%%%%%%%%%%%%%%%%%%%%%%%%%%%%%%%%%%%%%%%%%%%%%%%%%%%
to be compared with the value $\kappa = 8.25$ obtained
in Ref. \cite{Malek98} using the incremental expansion.
Note that the error in the determination of $\kappa$ is
only due to the limited precision of $\epsilon_4$.
Nevertheless, the fair agreement with the result of \cite{Malek98}
indicates that the finite-size scaling of our Lanczos data
yields reasonable results for $\epsilon_4$.

\begin{figure}[t]
\centerline{\includegraphics[height=6cm,width=9cm]{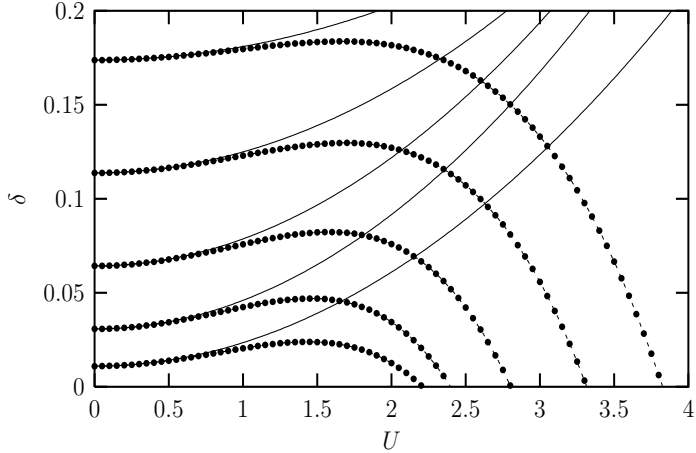}}
\caption{Dimerization $\delta$ as function of $U$ for
$g = 0.4, 0.45, 0.5, 0.55, 0.6$
(from bottom to top).
The full lines (circles) represent the second (fourth) order approximation of
$\delta(u) = \delta_0 + \delta_2 U^2 + \delta_4 U^4 + \ldots$. The
values of $U$ where $\delta$ becomes negative indicate the breakdown of the
weak coupling approximation.}
\label{fig:dofu}       % Give a unique label
\end{figure}

In Fig.\ \ref{fig:dofu} the equilibrium dimerization $\delta$
of the Peierls-Hubbard model is displayed as function of $U$ for
several values of the el-ph coupling $g$. The circles represent
the dimerization including the fourth order term in $U$
while the solid curves correspond to the second order approximation. For
small values of $g$ the maximum of $\delta$ is more pronounced
and occurs around $U \approx 2$. This is somewhat smaller than
the value $U \approx 3$ obtained using the incremental expansion \cite{Malek98}.
Furthermore the maximum of $\delta$ is expected to shift to larger
values of $U$ with decreasing $g$ but of course the precise position of the maximum
depends on the terms of order $\propto U^6$ and higher which have been
omitted in our weak coupling approach.
%%%%%%%%%%%%%%%%%%%%%%%%%%%%%%%%%%%%%%%%%%%%%%%%%%%%%%%%%%%%
%%%%%%%%%%%%%%%%%%%%%%%%%%%%%%%%%%%%%%%%%%%%%%%%%%%%%%%%%%%%%%%%%%%%%%%%%%%%
\section{Discussion}
%%%%%%%%%%%%%%%%%%%%%%%%%%%%%%%%%%%%%%%%%%%%%%%%%%%%%%%%%%%%%%%%%%%%%%%%%%%%
We have investigated the Peierls Hubbard model in the weak-coupling
regime using second order perturbation theory
and exact diagonalization for the fourth order contribution.
The application of perturbation theory is justified by the existence of
a gap in the excitation spectrum which removes the problem of
small denominators. On the other hand, it is also due to this gap that
finite-size scaling of the Lanczos data becomes problematic for
small dimerization $\delta < 0.2$ where the correlation length exceeds
the size of the systems that can be diagonalized numerically.
The central result of this paper is the determination of the exact value
of the el-ph coupling $g_c = 0.689348$ which separates the region
where the Hubbard interaction enhances the dimerization $(g < g_c)$
from the region where the dimerization is reduced $(g > g_c)$.
The corresponding dimerization parameter is found to be $\delta_c = 0.310523$.
There have been many attempts to determine these parameters using
different methods. Using the Gutzwiller variational wave-function
Baeriswyl and Maki \cite{Baeriswyl85} estimated $g_c \approx 0.75$.
Exact diagonalization of small systems has been
used  to obtain $\delta_c \approx 0.4$ in \cite{Hayden88} and
$g_c \approx 0.76$ in \cite{Waas90}, respectively.
Finally, the most precise values up to now have been derived using
the incremental expansion technique, which is based on the
numerical diagonalization of finite chains. In \cite{Malek98,Malek03}
$g_c = 0.69$ is obtained, which is very close to the exact value.

As mentioned before, all approaches that rely on finite-size extrapolation
of small systems suffer from severe problems in the limit of small dimerization,
$\delta < 0.1$, which corresponds to values of the el-ph coupling
$g < 0.5$.
In contrast, the perturbation theory approach can be applied for all values of
$g$. In particular, it is possible to determine the asymptotic
behavior of $\epsilon_2(\delta)$ in the limit $\delta \rightarrow 0$
including logarithmic corrections. From there
it can be inferred that the {\it relative} importance
of the second order term  (compared to the leading term) in the expansion
of the dimerization $\delta(U) = \delta_0 + \delta_2 U^2 + \delta_4 U^4 + \dots$
goes as $\sim g^{-6}$ for $g \rightarrow 0$, although both $\delta_0$ and $\delta_2$ are
exponentially small in this limit.
On the other hand, considering {\it absolute} values,
the correlation enhancement of the dimerization
is less dramatic and essentially limited to the region $0.3 < g < g_c$
with the maximum occurring at $g = 0.464$. The parameters most frequently used
for polyacetylene lie in the range between $g = 0.39$ \cite{Baeriswyl85} and
$g = 0.57$ \cite{Su80} which emphasizes the importance of correlation
effects in real Peierls systems.
%%%%%%%%%%%%%%%%%%%%%%%%%%%%%%%%%%%%%%%%%%%%%%%%%%%%%%%%%%%%%%%%%%%%%%%%%%%%
\section*{Acknowledgements}
We thank C\ Schuster, P\ Schwab, U\ Eckern, and H\ Zheng for helpful discussions.
This work was supported by the Deutsche Forschungsgemeinschaft (SFB 484).
%%%%%%%%%%%%%%%%%%%%%%%%%%%%%%%%%%%%%%%%%%%%%%%%%%%%%%%%%%%%%%%%%%%%%
%                        REFERENCES                                 %
%%%%%%%%%%%%%%%%%%%%%%%%%%%%%%%%%%%%%%%%%%%%%%%%%%%%%%%%%%%%%%%%%%%%%
%

%%%%%%%%%%%%%%%%%%%%%%%%%%%%%%%%%%%%%%%%%

\section*{References}
\begin{thebibliography}{99}

\bibitem{Peierls55}
Peierls R~E 1955
{\it Quantum Theory of Solids} (Oxford: Oxford University Press)

\bibitem{Keiss92}
Kiess H~G 1992
{\it Conjugated Conducting Polymers} (Berlin: Springer)

\bibitem{Ishiguro90}
Ishiguro T and Yamaji K 1990
{\it Organic Superconductors} (Berlin: Springer)

\bibitem{Su80}
Su W~P, Schrieffer J~R and  Heeger A~J 1979
{\it Phys.\ Rev.\ Lett.\ } {\bf 42} 1698

Su W~P, Schrieffer J~R and  Heeger A~J 1980
{\it Phys.\ Rev.\ B} {\bf 22} 2099

\bibitem{Sengupta03}
Sengupta P, Sandvik A~W and Campbell D~K 2003
{\it Phys.\ Rev.\ B} {\bf 67} 245103

\bibitem{Horsch81}
Horsch P 1981
{\it Phys.\ Rev.\ B} {\bf 24} 7351

\bibitem{Baeriswyl85}
Baeriswyl D and Maki K 1985
{\it Phys.\ Rev.\ B} {\bf 31} 6633

\bibitem{Kivelson82}
Kivelson S and Heim D~E 1982
{\it Phys.\ Rev.\ B} {\bf 26} 4278

\bibitem{Hirsch83}
Hirsch J~E 1983
{\it Phys.\ Rev.\ Lett.\ } {\bf 51} 296

\bibitem{Mazumdar83}
Mazumdar S and Dixit S~N 1983
{\it Phys.\ Rev.\ Lett.\ } {\bf 51} 292

\bibitem{Hayden88}
Hayden G~W and Soos Z~G 1988
{\it Phys.\ Rev.\ B} {\bf 38} 6075

\bibitem{Waas90}
Waas V, B\"uttner H and Voit J 1990
{\it Phys.\ Rev.\ B} {\bf 41} 9366

\bibitem{Ogata93}
Ogata M 1993
{\it Prog. Theor. Phys. Supp.} {\bf 113} 215

\bibitem{Sugiura02}
Sugiura M and Suzumura Y 2002
{\it J.\ Phys.\ Soc.\ Jpn.\ } {\bf 71} 697

\bibitem{Mocanu04}
Mocanu C, Dzierzawa M, Schwab P and Eckern U 2004
{\it J.\ Phys.\ : Condens.\ Matter} {\bf 16} 6445

Mocanu C, Dzierzawa M, Schwab P and Eckern U 2005
{\it Phys. Status Solidi b} {\bf 242} 245

\bibitem{Malek98}
M\'alek J, Kladko K and Flach S 1998
{\it JETP Lett.\ } {\bf 67} 1052

\bibitem{Malek03}
M\'alek J, Drechsler S-L, Flach S, Jeckelmann E and Kladko K 2003
{\it J.\ Phys.\ Soc.\ Jpn.\ } {\bf 72} 2277

\bibitem{Lieb68}
Lieb E~H and Wu F~Y 1968
{\it Phys.\ Rev.\ Lett.\ } {\bf 20} 1445

\bibitem{Economou79}
Economou E~N and Poulopoulos P~N 1979
{\it Phys.\ Rev.\ B} {\bf 20} 4756

\bibitem{Metzner89}
Metzner W and Vollhardt D 1989
{\it Phys.\ Rev.\ B} {\bf 39} 4462

\bibitem{Krivnov84}
Krivnov V~Ya and Ovchinnikov A~A 1984
{\it JETP Lett.} {\bf 39} 159;
1986 {\it Zh. Eksp. Teor. Fiz.} {\bf 90} 709

\bibitem{Horovitz85}
Horovitz B and S\'olyom J 1985
{\it Phys.\ Rev.\ B} {\bf 32} 2681

\bibitem{Kivelson86}
Kivelson S, Thacker H and Wu W-K 1985
{\it Phys.\ Rev.\ B} {\bf 31} 3785

\bibitem{Wu86}
Wu W-K and Kivelson S 1986,
{\it Phys.\ Rev.\ B} {\bf 33} 8546

\end{thebibliography}
\end{document}